\documentclass[runningheads]{llncs}
\usepackage{graphicx}
\usepackage{comment} % to comment out sections.
\usepackage{float} % to fix table positions
\usepackage{seqsplit} % to split long urls
\usepackage{subcaption} % support subfigures captions
\usepackage{longtable} % to split tables on multiple pages
\usepackage{fontawesome}

\usepackage{algorithm} 
\usepackage[noend]{algpseudocode} 
\MakeRobust{\Call} 
\algrenewcommand\algorithmicindent{0.2cm}

\usepackage{todonotes}

\begin{document}
\title{A Semi-automated Method for Domain-Specific Ontology Creation from Medical Guidelines}
\titlerunning{Domain-Specific Ontology Creation}

\author{Omar ElAssy\orcidID{0000-0001-6202-2465} \and
Rik de Vendt\orcidID{0000-0002-2131-0425} \and Fabiano Dalpiaz\orcidID{0000-0003-4480-3887} \and Sjaak Brinkkemper\,\textsuperscript{\faEnvelopeO}\ \orcidID{0000-0002-2977-8911}}
\authorrunning{O. ElAssy et al.}

\institute{Dept. of Information and Computing Sciences, Utrecht University, The Netherlands \\
\email{\{o.omarihabelsayedelassy, r.devendt, f.dalpiaz, s.brinkkemper\}@uu.nl}}
\maketitle  

\begin{abstract}
The automated capturing and summarization of medical consultations has the potential to reduce the administrative burden in healthcare. Consultations are structured conversations that broadly follow a guideline with a systematic examination of predefined observations and symptoms to diagnose and treat well-defined medical conditions. 
A key component in automated conversation summarization is the matching of the  knowledge graph of the consultation transcript with a medical domain ontology for the interpretation of the consultation conversation. Existing general medical ontologies such as SNOMED CT provide a taxonomic view on the terminology, but they do not capture the essence of the guidelines that define consultations.
As part of our research on medical conversation summarization, this paper puts forward a semi-automated method for generating an ontological representation of a medical guideline. The method, which takes as input the well-known SNOMED CT nomenclature and a medical guideline, maps the guidelines to a so-called Medical Guideline Ontology (MGO), a machine-processable version of the guideline that can be used for interpreting the conversation during a consultation.
We illustrate our approach by discussing the creation of an MGO of the medical condition of ear canal inflammation (Otitis Externa) given the corresponding guideline from a Dutch medical authority.
\end{abstract}

\section{Introduction}
The automated summarization of conversations may save time and cost, especially in domains where dialogues are structured based on predefined guidelines. Medical consultations are a prime example, as they broadly follow a systematic examination of predefined symptoms and observations to diagnose and treat well-defined conditions affecting fixed human anatomy \cite{latif2020speech}. The potential of automated conversation summarization depends on the readiness of the structured representations of domain-specific knowledge in a machine-processable format \cite{maas2020care2report}. In particular, rule-based conversation summarization grants explainability that is unachievable with machine learning approaches, thus, ontologies can efficiently represent domain guidelines in such rule-based applications \cite{wang2021systematic}. 

\begin{figure}[!h]
\centering
\includegraphics[width=1\textwidth]{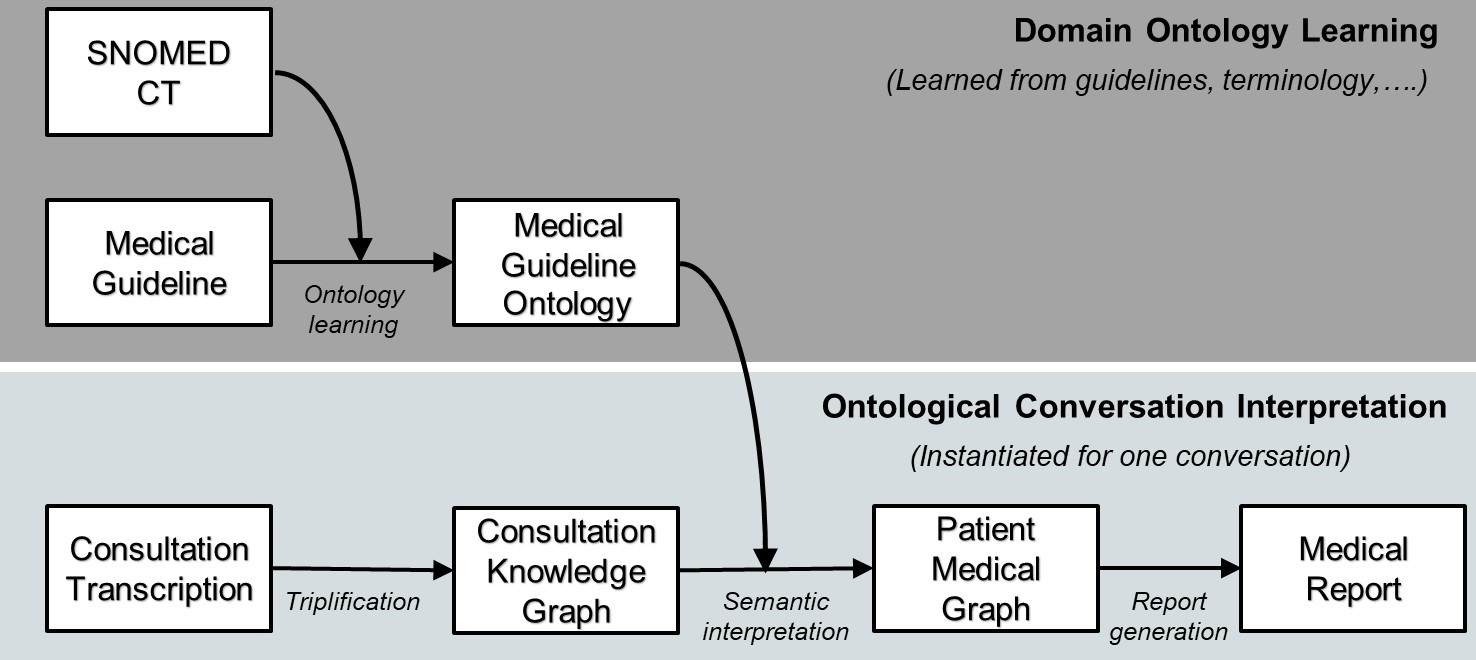}
\caption{Ontological Conversation Interpretation in Care2report~
\cite{molenaar2020medical}, with the Medical Guideline Ontology supporting semantic interpretation.}
\label{fig:c2rstages}
\end{figure}

Medical consultations typically last 5--10 minutes in which about 1,000--1,500 words are spoken by care provider and patient. Medical reports contain between 20 and 40 highly standardized terms with abbreviations for frequent words (e.g., pt for patient) \cite{molenaar2020medical}. In our Care2Report research project \cite{maas2020care2report,molenaar2020medical}, we combine \textit{domain ontology learning} with the concept of \textit{ontological conversation interpretation} in a two-layered architecture (Fig.~\ref{fig:c2rstages}).
The semantic interpretation of the consultation conversation requires the availability of a domain-specific ontology, that binds spoken words to formally defined medical concepts. Input are medical guidelines and clinical practice guidelines for the patient’s ailment type that are available from medical professional associations. Many ontology learning techniques (linguistic analysis, inductive logic programming, and statistical learning \cite{asim2018survey}) can be applied to generate a domain-specific ontology, in our case the Medical Guideline Ontology (MGO). As medical conversations require optimal precision, we prefer linguistic parsing analyzers (see Sec.~\ref{sec:method}). The SNOMED CT medical glossary \cite{snomed} serves for the medical concept identification and linking.

\vspace{-0.5cm}
The \textit{ontological conversation interpretation pipeline} enables the consultation summarization in four steps: (1) transcribing the audio file by utilizing an automated speech recognizer (ASR), (2) generation of OWL triples (aka triplificiation) into a complete consultation knowledge graph, (3) semantic interpretation of these triples by means of the Medical Guideline Ontology to a significantly smaller patient medical graph \cite{molenaar2020medical}, and (4) the generation of a medical report that is presented to the care provider for editing, approval and uploading into the Electronic medical file of this patient. 

We adopt the distinction of Ehrlinger and W{\"o}{\ss  ~\cite{ehrlinger2016towards}: the medical ontology acts as a schema on a generic layer serving multiple similar consultations, whereas the knowledge graph represents an instantiated fact base as spoken during the consultation. %So the ontology is the explicit specification of a conceptualization \cite{gruber1993translation}, a formal description of the concepts in this medical domain and the relationships between these concepts.  
In the context of conversation summarization, the ontology represents the diverse and primarily textual domain knowledge (e.g., human anatomy and medical guidelines) in a machine-processable fabric of concepts and relationships. Manually building a domain ontology is time consuming and error prone; thus, it should be carried out using ontology learning: constructing and integrating a machine-readable semantic-rich ontology (semi-)automatically  \cite{zhou2007ontology}.

However, tools and techniques for ontology learning primarily aim to extract knowledge from general data corpora as it was, and still is, driven by the necessity of linking either openly available data (e.g., DBpedia \cite{lehmann2015dbpedia}) or corporate-specific business data (e.g., Google Knowledge Graph \cite{singhal2012}). Therefore, research is needed to develop a method that enables the systematic representation of domain guidelines in machine-processable ontologies. This paper  introduces the notion of \textit{ontological conversation interpretation} (Fig.~\ref{fig:c2rstages}) by presenting a method for developing ontologies in the medical domain to support the expansion of the automated conversation summarization system Care2Report \cite{maas2020care2report}.  

Care2Report's summarization pipeline \cite{maas2020care2report} relies on an ontology embodying the domain's vocabulary and guidelines that act as the structured container to be filled with the multimodal information from the medical consultation.
The information extracted from the conversation populates the ontology to generate the rule-adhering report that the physician checks before uploading to the Electronic Medical Records system (EMR) \cite{molenaar2020medical}. This paper investigates elements of the second stage of the Care2Report pipeline: the definition of a Medical Guideline Ontology, a machine-processable version of a medical guideline. %The multimodal data recorded and processed during the medical consultation is formally expressed using a knowledge representation formalism to allow further analysis in later stages.

To illustrate, we use a medical guideline for ear canal inflammation (Otitis Externa) from the Dutch College of General Practitioners~\cite{nhgguidelines}. The medical domain knowledge is assembled from the terminology in the Systematized Nomenclature of Medicine - Clinical Terms (SNOMED CT)~\cite{snomed}. 

This paper addresses the two aspects of the ontology learning method: the notational side represented by the Medical Guideline Ontology (MGO), and the procedural method expressed in a Process-Deliverable Diagram (PDD) model and by an algorithm. The design science research cycle, as described by Wieringa \cite{wieringa2014design}, is fitting to answer the research question: \textit{How to systematically construct ontologies from the human anatomy and medical guidelines?}

The paper makes three contributions:
\begin{itemize}
    \item We introduce and formalize the Medical Guideline Ontology (MGO), a domain ontology that constitutes a machine-readable version of a medical guideline and that describes the relevant aspects concerning the patient's anatomy, symptoms, physician's observations, diagnosis, and treatments.
    \item We define a procedural method to develop such ontologies in the form of a Process-Deliverable Diagram model, refined into an algorithm that can be at the basis of automated tooling.
    \item We illustrate the MGO and the procedure to the case of the external ear canal inflammation.
\end{itemize}

\textit{Paper organization.} After reviewing the related work in Sec.~\ref{sec:bg}, we introduce the MGO in Sec.~\ref{sec:pmo} and its formalization in Sec.~\ref{sec:pmoformalized}. We then present our method in Sec.~\ref{sec:method} and its application in Sec.~\ref{sec:application}. Finally, we draw conclusions in Sec.~\ref{sec:concl}.

\section{Guidelines and Nomenclature in Medical Informatics}
\label{sec:bg}

\textbf{Medical Guidelines.} A medical guideline is a document with recommendations that aim to optimize patient care based on a systematic review of evidence and on weighing the benefits and harms of alternative care options \cite{peleg}. It consists of definitions and procedural instructions for executing an anamnesis, diagnosis and treatment in care provisions that aim to advance care quality, improve the patient’s health and well-being, and support medical decision-making. Many (inter)national medical authorities publish and maintain medical guidelines \cite{world2014handbook}. 

In the Netherlands, both the Dutch College of General Practitioners (Nederlands Huisartsen Genootschap - NHG) and the Dutch Federation of Medical Specialists (Federatie Medisch Specialisten - FMS) publish numerous guidelines  \cite{nhgguidelines,fmsguidelines}, only the former is used in this paper. The guidelines include sections about prognosis, common causes and background, physical examination and diagnosis guidelines, treatment policy, consultations and referral guidelines (if any), and control and future patient check-ups. 

The symptoms and observations indicating a condition and the treatments recommended for such condition by the guidelines are relevant to the construction and population of the Medical Guideline Ontology (MGO) and the related sub-ontologies, as will be detailed in the coming sections. The MGO, the consultation knowledge graph, and the consultation report aim to serve as a representational artifact, rather than supporting the physician’s decision making; thus, the reasoning behind which treatment to choose is beyond our scope. Therefore, the MGO should contain all treatment possibilities as options, while the physician’s discretion will decide which ones to use.

\vspace{0.2cm}
\noindent\textbf{Medical Nomenclature: SNOMED CT.}
The Systematized Nomenclature of Medicine - Clinical Terms (SNOMED CT) is currently the world’s most extensive clinical terminology system \cite{bodenreider2018recent}.  This paper uses SNOMED CT as an ontology source representing human anatomy and terminology hierarchy to identify relevant medical concepts from all the potential concepts in the textual medical guidelines. The terminology structure is in hierarchical formations of concepts defined and connected to each other by relationships, with identifiers for machine use and descriptors for human readability.
The top node of the SNOMED CT hierarchy is occupied by the root concept \textit{SNOMED CT concept}, and nineteen direct subtypes of it are the \textit{top level concepts} that provide the structure of the SNOMED CT. Different conditions and medical consultations use a subset of the available concept hierarchies.

% \begin{figure}[ht]
% \centering
% \includegraphics[width=1\textwidth]{snomed_design}
% \caption{SNOMED CT Design \cite{snomed}}
% \label{fig:snomed_design}
% \end{figure}

\section{Medical Guideline Ontology}
\label{sec:pmo}
The Medical Guideline Ontology (MGO) is a domain ontology that represents a medical guideline in a machine-processable format. In the context of the considered guideline, the MGO represents the relevant patient’s anatomy and symptoms, the physician’s observations, diagnosis and prescribed treatments. Note that symptoms are subjective abnormalities that the patient perceives, while observations are objective abnormalities detected by the physician  \cite{leblond2015degowin}. The schema of the MGO is illustrated in Fig.~\ref{fig:pmometa}.
\begin{figure}[!h]
\centering
\includegraphics[width=0.8\textwidth]{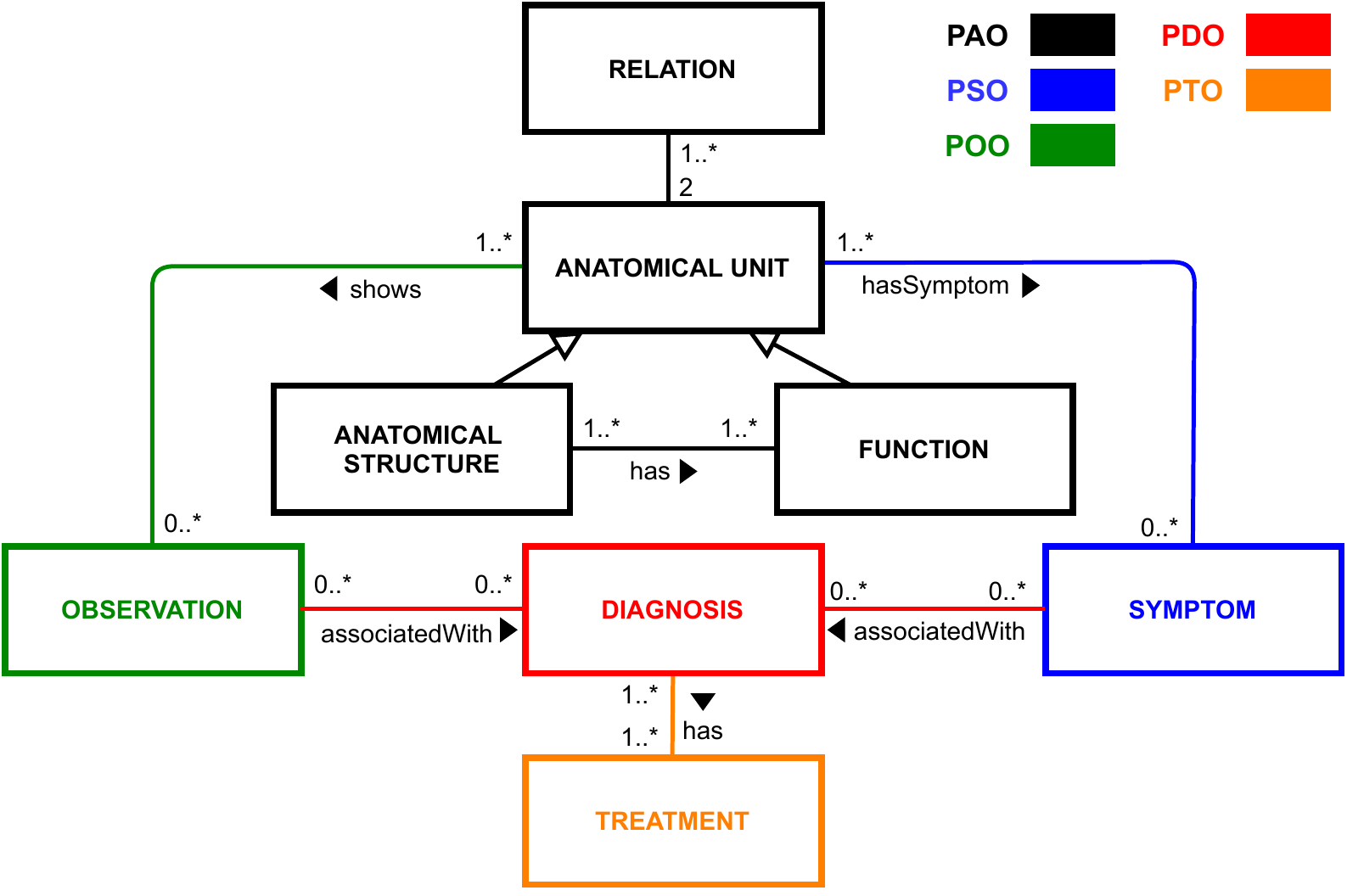}
\caption{Schema of the Medical Guideline Ontology}
\label{fig:pmometa}
\end{figure}

% \begin{figure}[!h]
% \centering
% \begin{subfigure}{1\textwidth}
% \centering
% \includegraphics[width=0.8\textwidth]{PMOmeta.pdf}
% \caption{Meta-model of the Medical Guideline Ontology}
% \label{fig:pmometa}
% \end{subfigure}\\
% \begin{subfigure}{1\textwidth}
% \centering
% \includegraphics[width=0.7\textwidth]{PMO}
% \caption{Medical Guideline Ontology reference notation}
% \label{fig:pmo}
% \end{subfigure}
% \caption{The Medical Guideline Ontology (PMO)}
% \label{fig:pmofull}
% \end{figure}

The MGO consists of five (sub)ontologies. The Patient Anatomy Ontology (PAO) depicts the human anatomical structures and functions (within the context of the guideline). The Patient Symptoms Ontology (PSO) represents the complaints and anomalies (symptoms) that a patient may report. The Patient Observations Ontology (POO) represents all the observations that the physician may make about the patient’s condition. The Patient Diagnosis Ontology (PDO) describes the physician’s diagnosis of a patient’s condition. Finally, the Patient Treatment Ontology (PTO) describes all the treatments prescribed by the physician, including medications, instructions, referrals, or additional medical tests.

The Patient Anatomy Ontology (PAO) is the foundation for the knowledge representation in the MGO. The PAO can be built based on existing resources like the Foundational Model of Anatomy \cite{rosse2003reference}, or it can utilize an existing hierarchical terminology structure like SNOMED CT, as in this research. Within the SNOMED CT hierarchies, the human anatomy is represented in one section: \textit{SNOMED CT concept $\Rightarrow$ Body structure $\Rightarrow$ Anatomical or acquired body structure} \cite{snomed}. The key concepts in the PAO are those of \textsc{Anatomical Unit} and \textsc{Relation}. The former is specialized by \textsc{Anatomical Unit} (e.g., ear, left eye) and \textsc{Function} (e.g., hearing), with the \textsc{has} relationship linking the two, e.g., `an ear has the function of hearing'. The \textsc{Relation} is meant to represent other types of links between anatomical units, such as part-of, next-to, etc.

The medical report produced by the automated conversation summarization system Care2Report to upload into the EMR follows a predefined structure \cite{cameron2002learning} of four sections: (i) Subjective: what the patient reports; (ii) Objective: what the physician identifies; (iii) Evaluation: assessment and diagnosis by the physician; and (iv) Plan for treatment and follow up.

Following the same arrangement, the MGO consists of a composite of four sub-ontologies reflecting the four aspects of reporting and a fifth foundational ontology for human anatomy.
Each of the those (anatomy, symptoms, observations, diagnosis, and treatment) builds on the previous one(s) to include further knowledge into the resulting MGO shown in Fig.~\ref{fig:pmometa}. The MGO provides the notational aspect of the ontology development method proposed in this paper.

The MGO and its instantiation into a consultation knowledge graph with information from a specific consultation (with a concrete patient) provide a graphical representation of ontology-based knowledge where every two concepts and the relationship between them form a triple. Examples of such triples are introduced in the coming sections.

\section{Formalization of the MGO}
\label{sec:pmoformalized}
The structure and the interaction between the various ontologies are relatively intuitive for simple conditions concerning comparatively simple body structures. However, adding the complete human anatomy and the guidelines from several medical authorities can make the ontologies less intuitive to understand and communicate. Therefore, it is essential to introduce formal descriptions of the ontologies and their components; this section introduces a few of these descriptions. Formally defining the MGO requires the definition of the following sets:

\begin{table}[!h]
\centering
\noindent\begin{tabular}{p{1cm} p{11cm}}
\multicolumn{1}{l}{Set} & \multicolumn{1}{l}{Description} \\  \hline
\textit{B} & anatomical structures of the body. \\
\textit{F} & anatomical functions of the body. \\ 
\textit{A} & anatomical units (\textit{A} = \textit{B} $\cup$ \textit{F}). \\ 
%\textit{P} & patients. \\%\\(A patient is also an anatomical unit: P $\subset$ A). 
\textit{S} & medical symptoms (reported by a patient). \\ 
\textit{O} & medical observations (observed by a physician). \\ 
\textit{V} & possible values to be assigned to symptoms \textit{s} $\in$ \textit{S} or observations \textit{o} $\in$ \textit{O}. \\ 
\textit{E} & explicit diagnoses. \\ 
\textit{I} & implicit diagnoses. \\ 
\textit{D} & medical diagnoses: $D=E\cup I$. \\ 
\textit{T} & medical treatments. \\
\end{tabular}
\label{tab:sets}
\vspace{-0.5cm}
\end{table}

\noindent
The elements in these sets may vary depending on the scope of the ontologies. Defining the comprehensiveness of these sets depends on the goal of the medical reporting and on the number and complexity o the guidelines. We leave the investigation of this aspect to future research.

The definitions of the ontologies comprising the MGO are defined with respect to a patient $p$ (the subject of the guideline, which is also the most comprehensive anatomical unit) as follows:
\begin{table}[H]
\centering
\begin{tabular}{p{.8cm} p{2cm} p{8.75cm}}
\multicolumn{1}{l}{Ontol.} &
  \multicolumn{1}{l}{Vertices} &
  \multicolumn{1}{l}{Edges} \\ \hline
PAO &
  \textit{A} = \textit{B} $\cup$ \textit{F} & 
  \vtop{
  \hbox{\strut $\{$(\textit{$b_1$}, \textit{$b_2$}) $|$ \textit{{$b_1$}},\textit{$b_2$} $\in$ \textit{B} $\wedge$ \textit{$b_1$} is a direct anatomical sub-part of \textit{$b_2$}\}} $\cup$ }
  \hbox{\strut (\{\textit{b}, \textit{f}) $|$ \textit{b} $\in$ \textit{B} $\wedge$ \textit{f} $\in$ \textit{F} $\wedge$ \textit{b} has an anatomical function \textit{f}$\}$} \\
PSO &
  \textit{A} $\cup$ \textit{S} $\cup$ \textit{V} &
  $\{$(\textit{a}, \textit{s}) , (\textit{s}, \textit{v}) $|$ \textit{a} $\in$ \textit{A} $\wedge$ \textit{s} $\in$ \textit{S} $\wedge$ \textit{v} $\in$ \textit{V}$\}$ i.e., the symptoms \\ 
POO &
  \textit{A} $\cup$ \textit{O} $\cup$ \textit{V} &
  $\{$(\textit{a}, \textit{o}) , (\textit{o}, \textit{v}) $|$ \textit{a} $\in$ \textit{A} $\wedge$ \textit{o} $\in$ \textit{O} $\wedge$ \textit{v} $\in$ \textit{V}$\}$ i.e., the observations \\ 
PDO &
  $\{p\}$ $\cup$ \textit{D} &
  $\{$(\textit{p}, \textit{d}) $|$ \textit{d} $\in$ \textit{D}$\}$ i.e., the possible diagnoses of a generic patient \\ 
PTO &
  $\{p\}$ $\cup$ \textit{T} &
  $\{$(\textit{p}, \textit{t}) $|$ \textit{t} $\in$ \textit{T}$\}$ i.e., the possible treatments of a generic patient
\end{tabular}
\label{tab:definitions}
\vspace{-0.5cm}
\end{table}
\noindent
As per Sec.~\ref{sec:pmo}, the MGO is a domain ontology representing the guideline's contents in terms of patient anatomy and symptoms as well as physician’s observations, diagnosis and treatments: MGO = PAO $\cup$ PSO $\cup$ POO $\cup$ PDO $\cup$  PTO.

Finally, some of the rules that need to be coded into the system to define the relationships between the entities are listed below, consulting domain experts is expected to add to this list. We provide some examples, but note that the creation of a comprehensive list of rules is domain-specific and goes beyond the purpose of this paper:
\begin{enumerate}
    \item Each anatomical structure (b) is a part of another anatomical structure (b) unless it is the complete body structure (b*).\\
    $\forall b_1 \exists b_2$ :  isPartOf($b_1,b_2$) \hfill  $b_1, b_2 \in$ \textit{B}, $\neg$ ($b_1$ = b*) \hspace{1cm} 
    
    \item Each function (f) is assigned to one or more anatomical structure (b).\\
    $\forall f \exists b$ :  hasFunction(b,f) \hfill $b \in$ \textit{B}, $f \in$ \textit{F} \hspace{1cm} 
    
    \item Each symptom (s) is associated with one or more anatomical unit (a).\\
    $\forall s \exists a$ :  hasSymptom(a,s) \hfill $s \in$ \textit{S}, $a \in$ \textit{A} \hspace{1cm} 
    
    \item Each observation (o) is associated with one or more anatomical unit (a).\\
    $\forall o \exists a$ :  hasObservation(a,o) \hfill $o \in$ \textit{O}, $a \in$ \textit{A} \hspace{1cm} 
    
    \item Patients are diagnosed with at least diagnosis (d).\\
    Given $p,\ \exists d$ :  diagnosedWith(p,d) \hfill $d \in$ \textit{D} \hspace{1cm}
    
    \item Patients are treated with a treatment (t). A treatment encompasses any physician’s prescription, including medications, instructions for the patient to follow, referral to a specialist, further tests, or any other procedure. \\
    Given $p,\ \exists t$ :  treatedWith(p,t) \hfill $t \in$ \textit{T} \hspace{1cm}
    
    \item An explicit diagnosis (e) is associated with a symptom (s) or an observation (o).\\
    $\forall e \exists s$ $\exists o$:  associatedWith(s,e) $\vee$ associatedWith(o,e)
    
    \hfill $e \in$ \textit{E}, $s \in$ \textit{S}, $o \in$ \textit{O} \hspace{1cm}
    
    \item An implicit diagnosis is not explicit. That is, an implicit diagnosis (i) is neither associated with a symptom (s) nor an observation (o). \\
    $\forall i \forall e$:  $\neg$ (i = e) \hfill $i \in$ \textit{I}, $e \in$ \textit{E} \hspace{1cm}

\end{enumerate}

\section{Method for Systematic Creation of Medical Ontologies}
\label{sec:method}
While the previous section introduced the MGO for representing a medical guideline in a machine-processable format, this section introduces the procedural method perspective outlined in a Process-Deliverable Diagram (PDD) model. The PDD illustrates the activities and artefacts of a specific process \cite{van2009meta}. 
%The model depicts the activities on the left-hand side of the PDD diagram using UML activity diagram notations, while the deliverables are on the right-hand side of the diagram using UML class diagram notations \cite{omg2003guide}. 
The model emphasizes the relationships between the activities and their deliverables by connecting them with dotted arrows across the diagram \cite{van2009meta}.

\subsection{Ontology Creation PDD}
Fig.~\ref{fig:pddhigh} shows the used PDD, in which the process is broken down into eight main activities. For simplicity, activities three to seven are illustrated (and manually performed) sequentially; however, they can also be executed in parallel.

\begin{figure}[!h]
\centering
\includegraphics[width=1\textwidth]{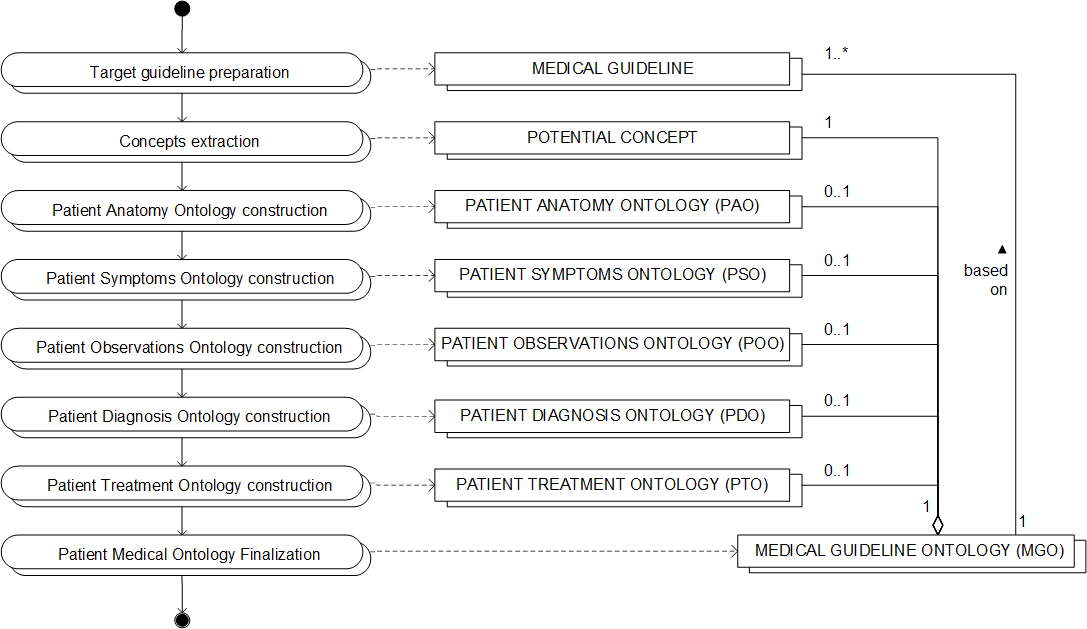}
\caption{PDD of the Ontology Creation Process}
\label{fig:pddhigh}
\end{figure}

\begin{enumerate}
    \item \textbf{Target guideline preparation}: The guideline of the medical condition is selected from the medical authority's website, translated (if necessary), scraped, and prepared for the following concept extraction activity. 
    \item \textbf{Concept extraction}: The relevant sections of the guidelines are identified, including sections describing symptoms, physical examination, and treatment plans. As nouns are the natural language representation of things, ideas and notions, they identify the concepts to be extracted. Thus, the potential concepts to extract are all nouns and noun phrases in the relevant sections that will be mapped in the next steps against the SNOMED CT to identify the constituent concepts of the ontology (e.g., anatomical units, symptoms, observations, and treatments). Some potential concepts will not be used in the ontology as they represent general nouns used in the text.
    \item \textbf{Patient Anatomy Ontology (PAO) construction}: The guideline concepts corresponding to SNOMED CT \textit{anatomical concepts} are identified and converted into a hierarchy from which the PAO is constructed.
    \item \textbf{Patient Symptoms Ontology (PSO) construction}: The concepts identified in the medical guideline sections describing symptoms are mapped against the corresponding SNOMED CT hierarchies (\textit{findings} and \textit{disorders}) to build the PSO.
    \item \textbf{Patient Observations Ontology (POO) construction}: Similar to the previous activity except dealing with physician observations instead of patient-described symptoms. Thus, the relevant guideline sections are different.
    \item \textbf{Patient Diagnosis Ontology (PDO) construction}: The medical condition or disease discussed in the guideline is the diagnosis associated with symptoms and observations to construct the PDO. 
    \item \textbf{Patient Treatment Ontology (PTO) construction}: The concepts identified in the treatment-related sections of the guideline are mapped against the corresponding SNOMED CT hierarchies (\textit{procedure}, \textit{substance}, \textit{dose form}, and \textit{physical object}) to build the PTO.
    \item \textbf{Medical Guideline Ontology Finalization}: All the previous (sub)ontologies are combined along with needed information (e.g., prefixes) to construct the complete MGO. This activity also includes checks to validity by confirming the lack of disjoint concepts.
\end{enumerate}

\subsection{A Detailed Look on Ontology Creation}
Fig.~\ref{fig:paopdd} expands the PAO construction activity to show the various subactivities and the resulting deliverables. 
\begin{figure} 
\centering
   \includegraphics[width=1\textwidth]{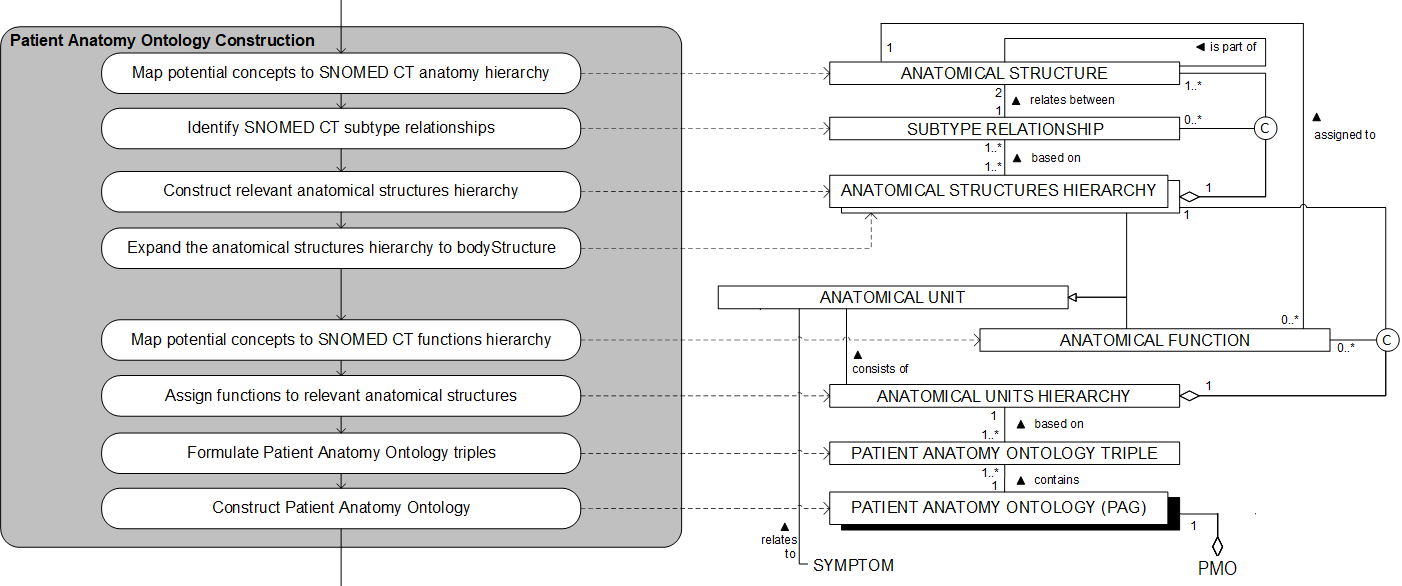}
   \caption{Patient Anatomy Ontology (PAO) construction}
   \label{fig:paopdd}
\end{figure}
The PDD activities constructing the remaining ontologies follow the same general high-level pattern to build the relevant ontology based on the list of extracted potential concepts from the guidelines. Potential concepts are mapped at each stage to both the relevant sections of the guidelines (e.g., physical examination and treatment policy) and the relevant hierarchies of the SNOMED CT (e.g., findings and disorders). Thus, the concepts that show in both relevant modules are the appropriate candidate concepts for the ontology at hand (e.g., symptom concept, observation concept or treatment concept). An anatomical unit (that can have symptoms, observations, diagnosis, or treatment) can be either an anatomical structure or a function. An \textit{anatomical structure} is ``a physical anatomical entity and a physical object, ... it consists of parts that are themselves anatomical structures", anatomical structures are ``...localized to a specific area or combine and carry out one or more specialized [anatomical] \textit{functions} of an organism." \cite{rosse1998motivation}.

For a detailed illustration of the complete PDD and a description of all the (sub)activities and concepts, refer to the technical report \cite{techrep}.

\subsection{Algorithm}
Algorithm~\ref{algo:ontogeneration} refines the PDD in Fig.~\ref{fig:pddhigh} by explaining the derivation of an MGO from a medical guideline $MG$ for a disease $D$, and from a human anatomy graph $\mathit{HA}$. This algorithm is at the basis of our current Care2Report pipeline. 
\begin{algorithm}[!h]
	\caption{Medical Guideline Ontology Generation}\label{algo:ontogeneration}
	
     \begin{flushleft}
     \textbf{Input}: $MG$ a Medical Guideline for disease $D$,\\
     \hspace*{\algorithmicindent}$\mathit{HA}$ the standard complete human anatomy graph,\\
     \textbf{Output}: a generic ontology $\mathit{MGO}$ for this medical guideline
    \end{flushleft}
	\begin{algorithmic}[1]
	\Function{BuildOntologyByTriples}{$MG$, $D$, $\mathit{HA}$}
	    \ForAll{$\mathit{sent} \in MG$}
	        \State $\mathit{P_{sent}}\gets\Call{ExtractNounPhrases}{sent}$ 
	        \ForAll{$\mathit{cs} \in \mathit{P_{sent}}$}
	          \If{$\mathit{cs} \in \mathit{HA}$}  
	            \State $\mathit{AnatD} \gets \mathit{AnatD} \cup \{cs\}$
	           \EndIf
	        \EndFor
	   \EndFor
	   \State $\mathit{PAO} \gets \Call{Subgraph}{AnatD, \mathit{HA}}$ \Comment The concepts in \textit{AnatD} and their links
	   \State $MGO\gets\emptyset$, $PSO\gets\emptyset$, $POO\gets\emptyset$, $PDO\gets\emptyset$, $PTO\gets\emptyset$, 
	   \ForAll{$\mathit{ae} \in \mathit{PAO}$}
	        \ForAll{$\mathit{sent} \in MG$}
	        \State $\mathit{P_{sent}}\gets\Call{ExtractNounPhrases}{sent}$ 
	        \ForAll{$\mathit{cs} \in \mathit{P_{sent}}$}
	          \If{$\mathit{cs} \in \mathtt{SNOMED.symptom}$}
	            \State $\mathit{PSO}\gets \mathit{PSO}\cup \{\langle \mathit{ae} , \mathit{hasSymptom} , \mathit{cs}\rangle,\ 	                          \langle \mathit{cs}, \mathit{associatedWith} , \mathit{D}\rangle \} $
	           \EndIf 
	          \If{$\mathit{cs} \in \mathtt{SNOMED.observation}$}
	            \State $\mathit{POO}\gets \mathit{POO}\cup \{\langle \mathit{ae}, \mathit{hasObservation}, \mathit{cs}\rangle,\ 
                  \langle\mathit{cs},\mathit{associatedWith}, \mathit{D}\rangle\}$ 
                \EndIf
	          \If{$\mathit{cs} \in \mathtt{SNOMED.diagnosis}$}
	            \State $\mathit{PDO}\gets \mathit{PDO}\cup \{\langle \mathit{patient}, \mathit{diagnosedWith}, \mathit{cs}\rangle\} $
	            \EndIf
	          \If{$\mathit{cs} \in \mathtt{SNOMED.treatment}$}
	            \State $\mathit{PTO}\gets \mathit{PTO}\cup \{\langle\mathit{patient}, \mathit{treatedWith}, \mathit{cs}\rangle,\ 	                          \langle \mathit{D}, \mathit{hasTreament}, \mathit{cs}\rangle\}$
	            \EndIf
	       \EndFor
	    \EndFor
	  \EndFor
	    \State $MGO\gets PAO \cup PSO \cup POO \cup PDO \cup PTO$ 
	    \State \Return $\mathit{MGO}$
	\EndFunction
	\end{algorithmic}
\end{algorithm}

The algorithms iterates over all sentences in the medical guideline (lines 2--6). Each sentence is parsed to extract the noun phrases (e.g., Otitis,  ear canal) that are mentioned in the text (line 3). These noun phrases are analyzed (lines 4--6) so that only those corresponding to elements of the human anatomy graph are retained and stored into the variable \textit{AnatD}.
The \textit{PAO} is defined (line 7) as the subgraph of the \textit{HA} that includes only the concepts in \textit{AnatD}, their descendants in the part-of hierarchy, and the relationships between the retained concepts.

After their initialization as empty sets (line 8), the remaining ontologies \textit{MGO, }\textit{PSO}, \textit{POO}, \textit{PDO}, \textit{PTO} are populated in lines 9--21.
The elements added to the \textit{PAO} are iterated over (lines 9--20): the sub-cycle starting in line 10 iterates again over the medical guidelines, which are again parsed by extracting the noun phrases (line 11).
Each of these noun phrases contributes to populating the various ontologies, depending on whether the noun phrase is a symptom (lines 13--14), observation (lines 15--16), diagnosis (lines 17--18), or treatment (lines 19--20). Finally, the MGO is defined (line 21) as the union of the five ontologies.

\section{Application to the Otitis Externa case}
\label{sec:application}
To illustrate the MGO and apply the derived procedure to create an ontology, we use the NHG guideline for the inflammation of the external ear canal (Otitis Externa) \cite{otitis}. The condition is chosen as an example for the relative simplicity of the associated guidelines and the lack of complicated medical procedures, terminology, or differential diagnosis. Otitis Externa is an inflammation caused by a disturbance in the acidic environment of the ear canal and it is usually associated with swimming. The symptoms reported by the patient may include ear pain, ear itching, fluid drainage from the ear, and hearing loss. In addition, the physician examines both the complaint-free ear and the affected ear for signs of scars, swelling, flaking, redness, the state of the eardrum, etc. \cite{otitis}. 

Some symptoms (e.g., hearing loss) are not required for an Otitis Externa diagnosis, indicating that while the MGO representation of the condition should include it as a possible symptom, some specific consultations might have this symptom present while others do not. Also, the guidelines indicate that the physician should check the eardrum; however, Otitis Externa is not associated with any observations regarding the eardrum, suggesting that any observation of the eardrum (e.g., rapture) might indicate a different diagnosis. As for the treatment, the guideline recommends that the physician should instruct the patient on how to clean the infected ear properly, and prescribe ear drops. Also, referral to a specialist is recommended if the condition does not improve promptly or if the patient is from a specific vulnerable group (e.g., elderly and diabetic). 

As an example, we consider a fictitious consultation for an Otitis Externa patient suffering from ear pain (no indication of itching, fluid drainage, or hearing loss), and the examination shows swelling, redness, and skin flaking in the external ear canal. Figure (\ref{fig:pmootitis}) represents the general MGO of the Otitis Externa condition with indication of the guideline sections forming it, while figure (\ref{fig:kgpatient}) is the consultation-specific knowledge graph for the fictitious patient.

\begin{figure}[!h]
\centering

\begin{subfigure}[H]{1\textwidth}
\centering
\includegraphics[width=0.9\textwidth]{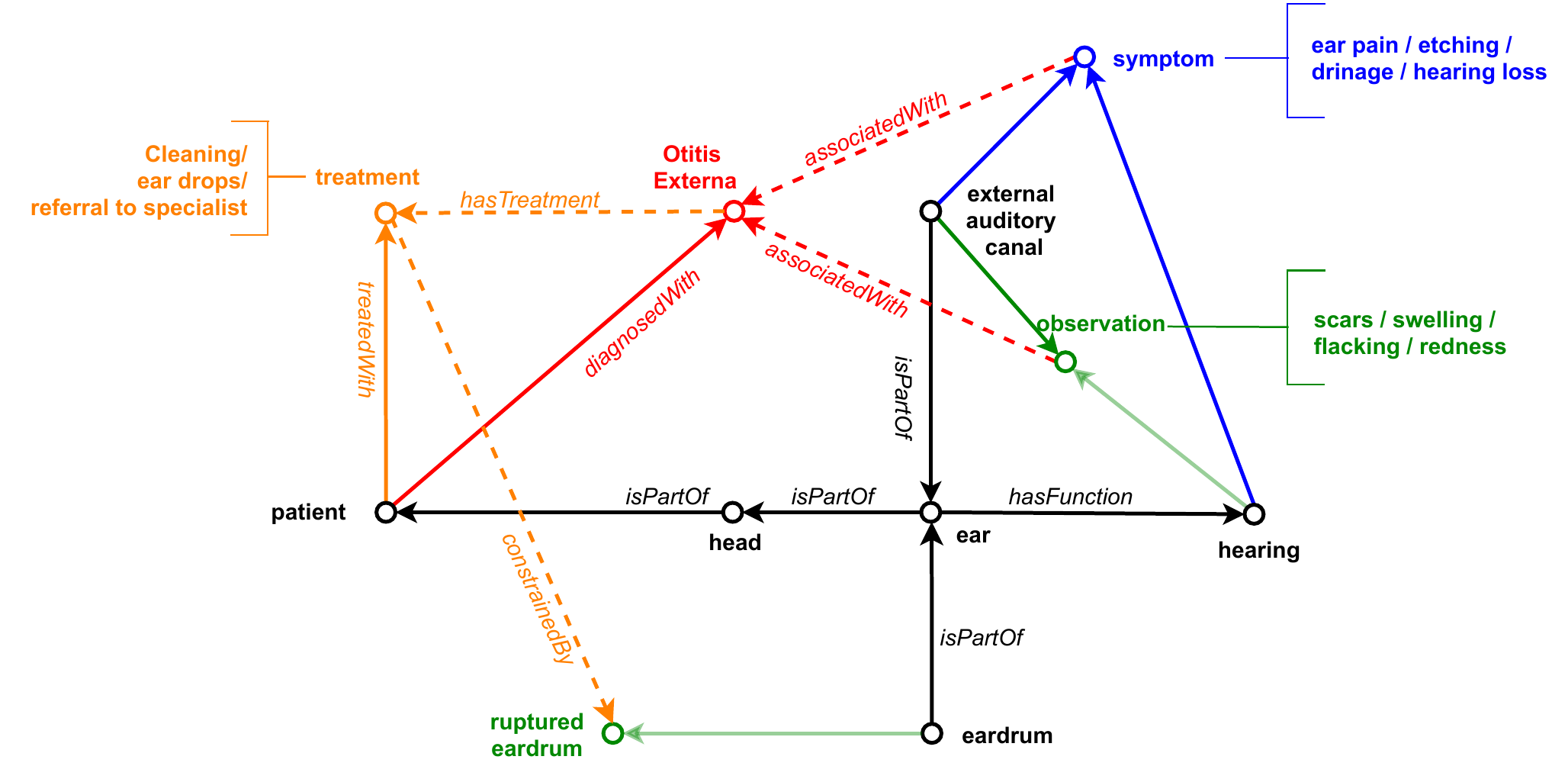}
\caption{MGO of the Otitis Externa Condition \cite{otitis}}
\label{fig:pmootitis}
\vspace{.3cm}
\end{subfigure}

\begin{subfigure}[H]{1\textwidth}
\centering
 \includegraphics[width=0.9\textwidth]{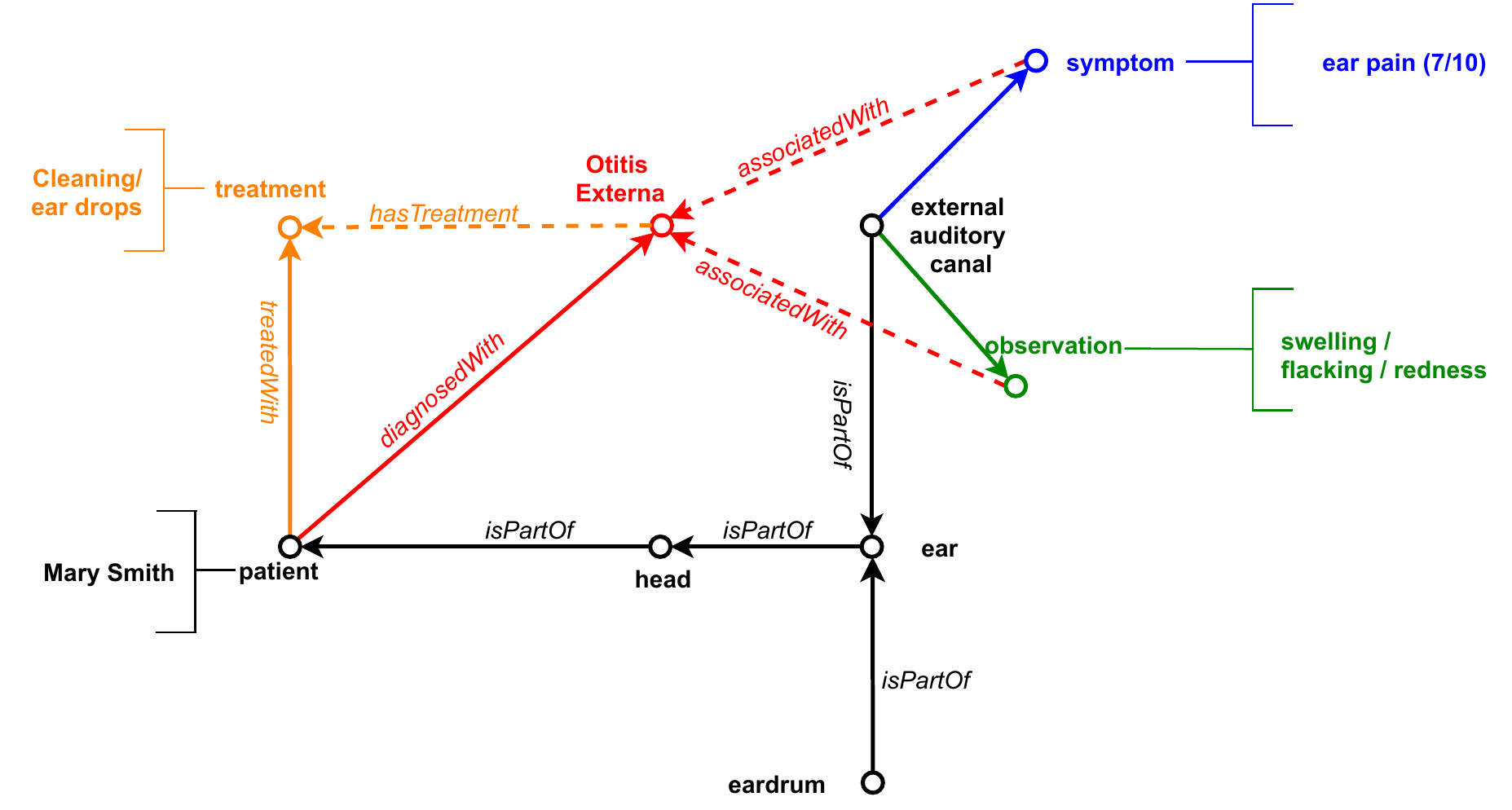}
 \caption{Example Consultation Knowledge Graph of an Otitis Externa patient}
 \label{fig:kgpatient}
\end{subfigure}
\caption{Otitis Externa representation}
\label{fig:fullotitis}
\vspace{-.3cm}
\end{figure}

The MGO can be represented in triples for machine processing; for example, some of the triples portrayed in figure (\ref{fig:kgpatient}) are explained below:

\begin{itemize}
    \item The human anatomy is represented in a hierarchy of anatomical structures connected to their parent structures using \textit{isPartOf} relationsips. For example, $\langle$externalAuditoryCanal, isPartOf, ear$\rangle$.
    \item The second part of the PAO is the assignment of anatomical functions to the anatomical structures performing them as in  $\langle$ear, hasFunction, hearing$\rangle$.
    \item Symtoms and observations triples define the PSO and the POO. For example, the following triples refer to a pain symptom: $\langle$externalAuditoryCanal, hasSymptom, symp\textunderscore2$\rangle$,  $\langle$symp\textunderscore2, symptom, earPain$\rangle$,  $\langle$symp\textunderscore2, hasValue, 7/10$\rangle$.
    \item The patient diagnosis and treatments are expressed in triples; for example: $\langle$patient, diagnosedWith, OtitisExterna$\rangle$, $\langle$patient, treatedWith, earDrops$\rangle$.
\end{itemize}

Applying the procedure in the PDD to Otitis Externa produces the ontology shown in Fig.~\ref{fig:OtitisWebVOWL} using the WebVOWL web application \cite{lohmann2016visualizing}. For conciseness, the ontology does not visualize the top-level classes (e.g., anatomy, symptom) and the is-a relationship linking the other concepts to them. More details on ontology construction and the resulting ontology can be found in the technical report \cite{techrep}.

\begin{figure}[!h]
\centering
   \includegraphics[width=\textwidth]{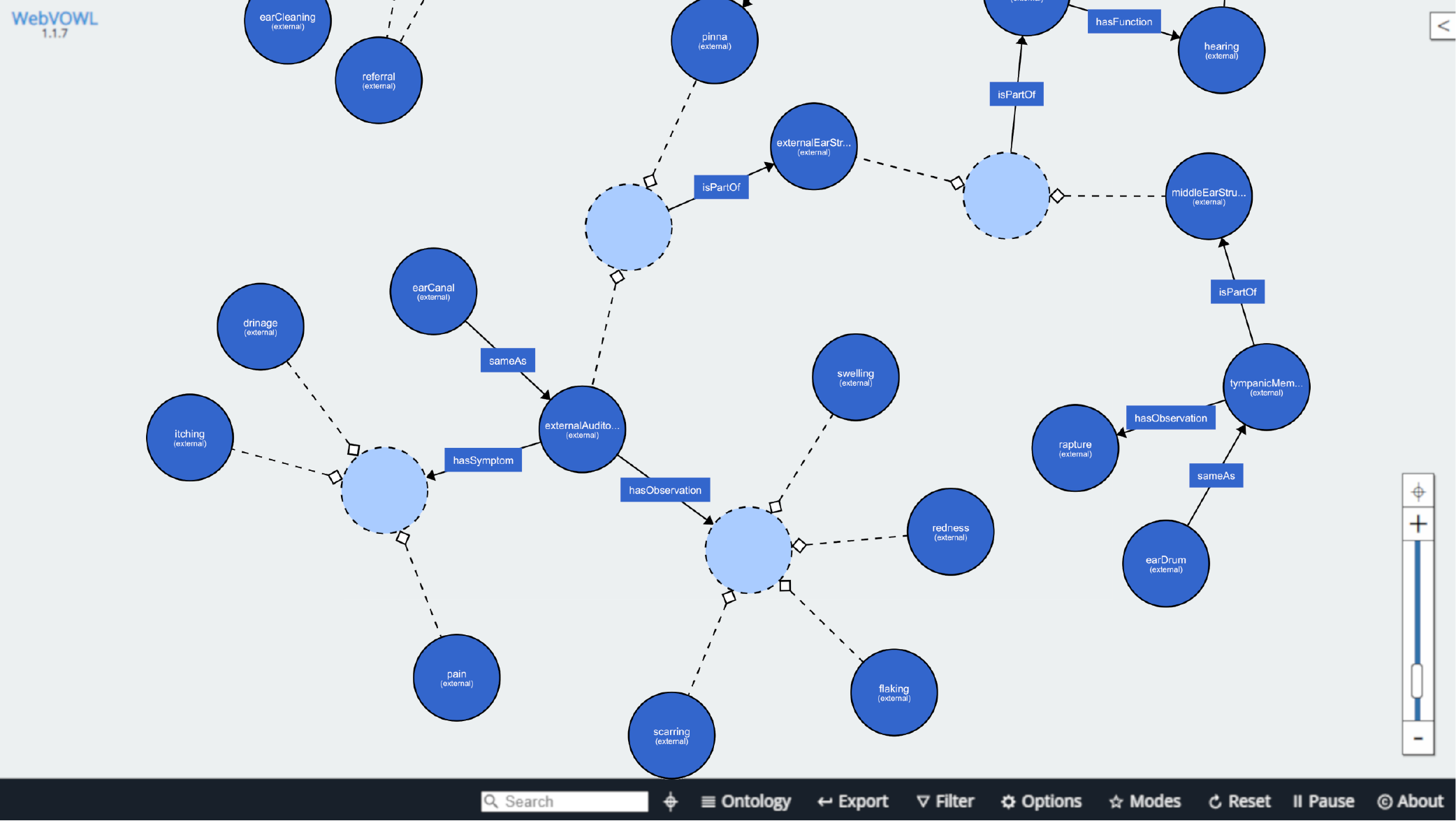}
   \caption{Partial WebVOWL Visualization of the Otitis Externa Ontology}
   \label{fig:OtitisWebVOWL}
\end{figure}

The method was applied manually, requiring substantial time investment. The full potential of the Care2Report system relies on developing an automatic process to generate ontologies of all medical conditions using guidelines from various countries and in different languages. The ontology development relies on data from two sources: SNOMED CT, which can be easily mapped to an OWL ontology; and the text-based medical guidelines that pose the real challenge for knowledge extraction and structuring to be tackled in future research.

\section{Conclusion}
\label{sec:concl}
We introduced \textit{ontological conversation interpretation} by creating a method for the systematic creation of medical ontologies from the SNOMED CT terminology and anatomical hierarchies, combined with the medical guidelines of different medical conditions. The resulting MGO is a formalized notation, and the PDD is a procedural guide to systematically create ontologies and enrich them by connecting new guidelines to the existing definitions of anatomy, symptoms, observations, and treatments; and adding more if the existing concepts are insufficient to represent the guideline fully. Furthermore, we have applied our approach to the case of Otitis Externa to illustrate its application and feasibility.

\paragraph{Limitations and Future Directions.} The method needs to be validated on guidelines of more complex and varying conditions. For example, the NHG guideline for ``Non-traumatic knee complaints" details a method for differentiating between various similar conditions and thus does not follow the typical sections in most NHG guidelines. This may require an evolution of our MGO. 
Moreover, some guidelines only point the physician towards the tests and measurements to monitor without detailing the results or values to look for, presumably because the results are hard to detail in the text while the medical professionals understand them. For example, the guidelines advise performing an electrocardiogram test in several cardiovascular conditions without detailing the expected outcomes. This is a challenge as the source information is incomplete; thus, the degree of possible automation is limited. The Care2Report multi-modal input architecture \cite{maas2020care2report} aims to eventually allow the integration of (some of) the measurement data, but an ontology to represent this knowledge still needs to be developed. 
Automation is another research challenge. We have applied the method manually, as explained in Sec.~\ref{sec:application}, and the correct and complete automated interpretation of textual medical guidelines is a far-fetched goal \cite{peleg}. Research is necessary toward the creation of assisted, interactive methods that support our approach. 
This includes research towards automating the creation of ontologies of the human anatomy and medical guidelines from different medical authorities.

This research aims to foster research in a societally-relevant field: increasing the quality of healthcare via semi-automated methods that may relieve medical professionals from their administrative burden. We make a step in this direction by laying down the formal foundations for the construction of semi-automated systems that support the interpretation of conversations using ontologies. 

\bibliographystyle{plain}
\bibliography{biblio}

\begin{thebibliography}{10}

\bibitem{fmsguidelines}
{Federatie Medisch Specialisten FMS - Richtlijnen (Guidelines of The Dutch
  Federation of Medical Specialists)}.
\newblock \url{https://richtlijnendatabase.nl/}.
\newblock Last accessed 11 March 2022.

\bibitem{singhal2012}
Introducing the knowledge graph: things, not strings.
\newblock
  \url{https://blog.google/products/search/introducing-knowledge-graph-things-not/}.
\newblock Last accessed 11 March 2022.

\bibitem{nhgguidelines}
{Nederlands Huisartsen Genootschap NHG - Richtlijnen (Guidelines of The Dutch
  College of General Practitioners)}.
\newblock \url{https://richtlijnen.nhg.org/}.
\newblock Last accessed 11 March 2022.

\bibitem{otitis}
{NHG Otitis Externa Guidelines}.
\newblock \url{https://richtlijnen.nhg.org/standaarden/otitis-externa}.
\newblock Last accessed 11 March 2022.

\bibitem{snomed}
{SNOMED CT Basics}.
\newblock
  \url{\seqsplit{https://confluence.ihtsdotools.org/display/DOCSTART/4.+SNOMED+CT+Basics}}.
\newblock Last accessed 11 March 2022.

\bibitem{asim2018survey}
Muhammad~Nabeel Asim, Muhammad Wasim, Muhammad Usman~Ghani Khan, Waqar Mahmood,
  and Hafiza~Mahnoor Abbasi.
\newblock A survey of ontology learning techniques and applications.
\newblock {\em Database}, 2018, 2018.

\bibitem{bodenreider2018recent}
Oliver Bodenreider, Ronald Cornet, and Daniel~J Vreeman.
\newblock Recent developments in clinical terminologies: {SNOMED CT, LOINC, and
  RxNorm}.
\newblock {\em Yearbook of Medical Informatics}, 27(01):129--139, 2018.

\bibitem{cameron2002learning}
Susan Cameron and Imani Turtle-Song.
\newblock Learning to write case notes using the {SOAP} format.
\newblock {\em Journal of Counseling \& Development}, 80(3):286--292, 2002.

\bibitem{ehrlinger2016towards}
Lisa Ehrlinger and Wolfram W{\"o}{\ss}.
\newblock Towards a definition of knowledge graphs.
\newblock {\em SEMANTiCS (Posters, Demos, SuCCESS)}, 48(1-4):2, 2016.

\bibitem{techrep}
Omar ElAssy, Fabiano Dalpiaz, and Sjaak Brinkkemper.
\newblock {\em {Developing Ontologies of Medical Guidelines for Automated
  Conversation Summarization}}, April 2022.
\newblock https://doi.org/10.5281/zenodo.6469617.

\bibitem{latif2020speech}
Siddique Latif, Junaid Qadir, Adnan Qayyum, Muhammad Usama, and Shahzad Younis.
\newblock Speech technology for healthcare: Opportunities, challenges, and
  state of the art.
\newblock {\em IEEE Reviews in Biomedical Engineering}, 14:342--356, 2020.

\bibitem{leblond2015degowin}
Richard~F LeBlond et~al.
\newblock {\em DeGowin's diagnostic examination}.
\newblock McGraw-Hill Education New York, 2015.

\bibitem{lehmann2015dbpedia}
Jens Lehmann, Robert Isele, Max Jakob, Anja Jentzsch, Dimitris Kontokostas,
  Pablo~N Mendes, Sebastian Hellmann, Mohamed Morsey, Patrick Van~Kleef,
  S{\"o}ren Auer, et~al.
\newblock Dbpedia--a large-scale, multilingual knowledge base extracted from
  wikipedia.
\newblock {\em Semantic Web}, 6(2):167--195, 2015.

\bibitem{lohmann2016visualizing}
Steffen Lohmann, Stefan Negru, Florian Haag, and Thomas Ertl.
\newblock Visualizing ontologies with vowl.
\newblock {\em Semantic Web}, 7(4):399--419, 2016.

\bibitem{maas2020care2report}
Lientje Maas, Mathan Geurtsen, Florian Nouwt, Stefan~F Schouten, Robin Van
  De~Water, Sandra Van~Dulmen, Fabiano Dalpiaz, Kees Van~Deemter, and Sjaak
  Brinkkemper.
\newblock The {Care2Report} system: Automated medical reporting as an
  integrated solution to reduce administrative burden in healthcare.
\newblock In {\em Proc. of HICSS}. 2020.

\bibitem{molenaar2020medical}
Sabine Molenaar, Lientje Maas, Ver{\'o}nica Burriel, Fabiano Dalpiaz, and Sjaak
  Brinkkemper.
\newblock Medical dialogue summarization for automated reporting in healthcare.
\newblock In {\em Proc. of EMMSAD}, pages 76--88. Springer, 2020.

\bibitem{world2014handbook}
World~Health Organization.
\newblock {\em WHO handbook for guideline development}.
\newblock World Health Organization, 2014.

\bibitem{peleg}
Mor Peleg.
\newblock Computer-interpretable clinical guidelines: A methodological review.
\newblock {\em Journal of Biomedical Informatics}, 46(4):744--763, 2013.

\bibitem{rosse1998motivation}
C~Rosse, JL~Mejino, BR~Modayur, R~Jakobovits, KP~Hinshaw, and JF~Brinkley.
\newblock Motivation and organizational principles for the digital anatomist
  symbolic knowledge base: An approach toward standards in anatomical knowledge
  representation.
\newblock {\em Journal of the American Medical Informatics Association},
  5:17--40, 1998.

\bibitem{rosse2003reference}
Cornelius Rosse and Jos{\'e}~LV Mejino~Jr.
\newblock A reference ontology for biomedical informatics: {The Foundational
  Model of Anatomy}.
\newblock {\em Journal of Biomedical Informatics}, 36(6):478--500, 2003.

\bibitem{van2009meta}
Inge van~de Weerd and Sjaak Brinkkemper.
\newblock Meta-modeling for situational analysis and design methods.
\newblock In {\em Handbook of research on modern systems analysis and design
  technologies and applications}, pages 35--54. IGI Global, 2009.

\bibitem{wang2021systematic}
Mengqian Wang, Manhua Wang, Fei Yu, Yue Yang, Jennifer Walker, and Javed
  Mostafa.
\newblock A systematic review of automatic text summarization for biomedical
  literature and {EHRs}.
\newblock {\em Journal of the American Medical Informatics Association},
  28(10):2287--2297, 2021.

\bibitem{wieringa2014design}
Roel~J Wieringa.
\newblock {\em Design science methodology for information systems and software
  engineering}.
\newblock Springer, 2014.

\bibitem{zhou2007ontology}
Lina Zhou.
\newblock Ontology learning: state of the art and open issues.
\newblock {\em Information Technology and Management}, 8(3):241--252, 2007.

\end{thebibliography}

%\begin{thebibliography}{8}

% \bibitem{ajami2016use}
% Ajami, S.: Use of speech-to-text technology for documentation by healthcare providers. The National medical journal of India \textbf{29}(3), 148 (2016)

% \bibitem{arndt2017tethered}
% Arndt, B.G. et al.: Tethered to the EHR: primary care physician workload assessment using EHR event log data and time-motion observations. The Annals of Family Medicine \textbf{15}(5), 419–426 (2017)

% \bibitem{babbott2014electronic}
% Babbott, S. et al.: Electronic medical records and physician stress in primary care: results from the MEMO Study. Journal of the American Medical Informatics Association \textbf{21}(e1), e100–e106 (2014)

% \bibitem{care2reportsite}
% Care2Report - Mission, \url{https://sites.google.com/view/care2report/mission}. Last accessed 27 June 2021

% \bibitem{chiu2017speech}
% Chiu, C.-C. et al.: Speech recognition for medical conversations.arXiv preprint arXiv:1711.07274 (2017)

% \bibitem{friedberg2014factors}
% Friedberg, M.W. et al.: Factors affecting physician professional satisfaction and their implications for patient care, health systems, and health policy. Rand health quarterly \textbf{3}(4), (2014)

% \end{thebibliography}

\end{document}